# Formation of a *p-n* junction on the GaAs-surface by an Ar$^+$ ion beam

*V.M. Mikoushkin*[1], *V.V. Bryzgalov*[1], *S.Yu. Nikonov*[1],
*A.P. Solonitsyna*[1], *D.E. Marchenko*[2,3],

[1] *Ioffe Institute, 194021 Saint-Petersburg, Russia*
[2] *Technische Universität Dresden, D-01062 Dresden, Germany*
[3] *Helmholtz-Zentrum BESSY II, German-Russian Laboratory, D-12489 Berlin, Germany*

**Abstract.** The core-level and valence band electronic structure of the well-defined near-surface layer of n-GaAs (100) has been studied by synchrotron-based high-resolution photoelectron spectroscopy before and after modification of the layer by an Ar$^+$ ion beam in the 1.5 - 2.5 keV energy range. Conversion of the conductivity type from *n* into *p* has been revealed in the irradiated layer several nm thick. The effect manifests itself in shifts of the core-levels and valence band edge by the value comparable to the bandgap width. Transformation on the conductivity type has been assumed to be caused by Ga-antisite point defects generated by ion bombardment. The possibility of local formation of a *p-n* nanojunction within the ion-beam spot has been shown.

## I. INTRODUCTION

GaAs and GaAs-based semiconductors have played an important role in modern high-frequency electronics due to their high electron mobility [1,2]. Therefore development of different methods for preparation of the semiconductor surfaces for further technological stages of device fabrication has been of great importance. Chemical etching and mechanopolishing are the widely used among these methods [1,3-5]. They are exploited for etching of the native oxides, surface passivation, and reconstruction of the surface atomic structure [1]. Treatment of the GaAs surface by chemically neutral argon projectiles has been also widely used for decades in research and development of this material. For example, etching of the GaAs surface by low- and mid-energy Ar$^+$ ion beams is one of the most exploited methods for removing native oxides and surface contaminations in order to prepare an atomically clean surface for both further high vacuum treatments and studying the GaAs-based semiconductors by surface-sensitive methods including different kinds of electron spectroscopy [6,7]. Chemical neutrality ensures pure mechanical action of accelerated argon atoms on the material. An important advantage of the argon projectiles is that argon atoms do not remain in materials after thermalization. They escape the irradiated layer due to diffusion. Ar$^+$ ion beams have been shown to be a sophisticated tool for nano-structuring of the GaAs surface resulting in creation of a dot-like relief [8-10]. The Ar$^+$ ion beam was utilized also for homogenization of the GaAs near-surface nitrided layer by its mixing [11]. Development of the Ar$^+$ ion techniques has been scientifically supported by research of the structure and elemental composition of layers irradiated and modified by Ar$^+$ ion beams. In particular, the effect of the irradiated layer enrichment with gallium due to preferential sputtering of arsenic atoms has been revealed and studied in the wide energy range: from 100 eV to 100 keV [12-16]. In this paper, we show how the Ga-enrichment effect modifies the electronic structure of the irradiated layer and results in creation of the *p-n* junction in the near-surface nanolayer of GaAs.

## II. EXPERIMENTAL DETAILS

Two sets of experiments were carried out in different times in ultrahigh vacuum at the Russian–German synchrotron radiation beamline of the BESSY-II electron storage ring (Berlin) [17] by using two end-stations: the photoelectron spectrometer ("Mustang") equipped with hemispherical analyzer SPECS Phoibos 150 (SPECS GmbH) and the spectrometer with hemispherical analyzer CLAM-4 (VG). The data obtained with the latter spectrometer are specially noted in the text below. The photoelectron (photoemission) spectra were measured at the normal direction from the sample irradiated by a monochromatic x-ray photon beam at an angle θ = 55°. The spectra were measured at photon energies from the range $hv$ = 135 - 800 eV providing different information depths. The photon energy scale of the plane-grating monochromator (RG-PGM) was regularly calibrated using the Au $4f_{7/2}$ line of gold in order to maintain the accuracy of the binding energy determination better than ± 0.05 eV. The analyzer operating mode was chosen for the major part of measurements so as to provide high sensitivity and statistics, though with sufficiently high energy resolution of ΔE < 300 meV. This resolution was sufficient to study the relatively broad 3d lines of Ga and As characterized by large spin-orbit splitting exceeding the resolution. The better resolution ΔE ~ 150 meV was also used when it was needed, for example in determining the valence band edge position.

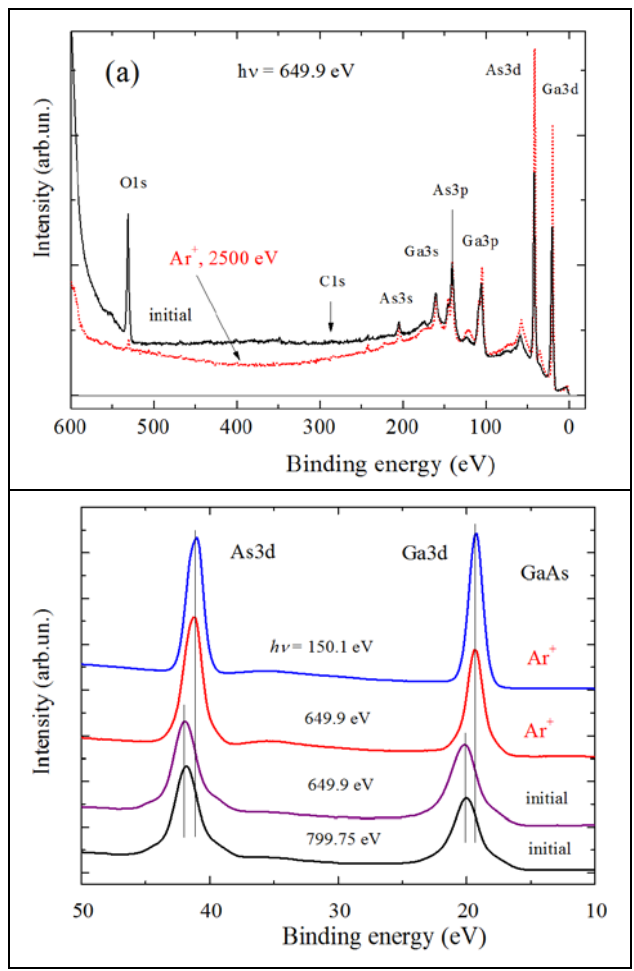

**Fig. 1.** Photoelectron spectra of a pristine n-GaAs (100) wafer covered by a layer of native oxide (initial) and of the same after removal of the oxide layer by the beam of $Ar^+$ ions with energy $E_i$ = 2500 eV ($Ar^+$): **(a)** survey spectra measured in a wide energy range, **(b)** As3d and Ga3d spectra normalized to the area of the As 3d line; the photon energy ($hv$) are indicated above the corresponding curves.

A commercial GaAs (100) n-type (n ~ $10^{18}$ $cm^{-3}$) wafer was studied before and after strong irradiation by $Ar^+$ ions with energies $E_i$ = 2500 eV and $E_i$ = 1500 eV directly in the preparation chamber of the spectrometer vacuum system. The survey spectra of the sample measured in a wide binding energy range are represented in Fig. 1a. The initial spectrum (initial - solid line) was measured from the sample slightly irradiated at the minimal dose density D < $1*10^{14}$ ions/$cm^2$ sufficient only to remove carbon contaminations from the surface. After etching the GaAs surface by Ar$^+$ ions with the dose density D ~ $3*10^{15}$ ions/cm$^2$ (Ar$^+$ - dotted line), the spectrum exhibits virtually absence of the O1s and C1s lines as well as of the lines of possible impurities, which evidences removal of the native oxide layer. The dose density D ~ $10^{16}$ Q/cm$^{-2}$ was sufficient to completely remove the layer of native oxide and a part of the bulk layer in the stationary sputtering regime. Thickness <x> of the irradiated layer was estimated as the width at the half maximum of the concentration depth distribution of the implanted projectiles calculated using the commonly known SRIM 2006 code [18]. The depth profile for the ion energy $E_i$ = 2500 eV is shown in Fig. 2a. The thicknesses of the irradiated layers were <x> ~ 7.5 and 5.4 nm for ion energies $E_i$ = 2500 eV and $E_i$ = 1500 eV, correspondingly. The maximum of the depth distribution corresponds to the projected range $R_p$ which is approximately equal to a half of the thickness ($R_p$ ~ ½<x>) and can be obtained by SRIM code with more ease.

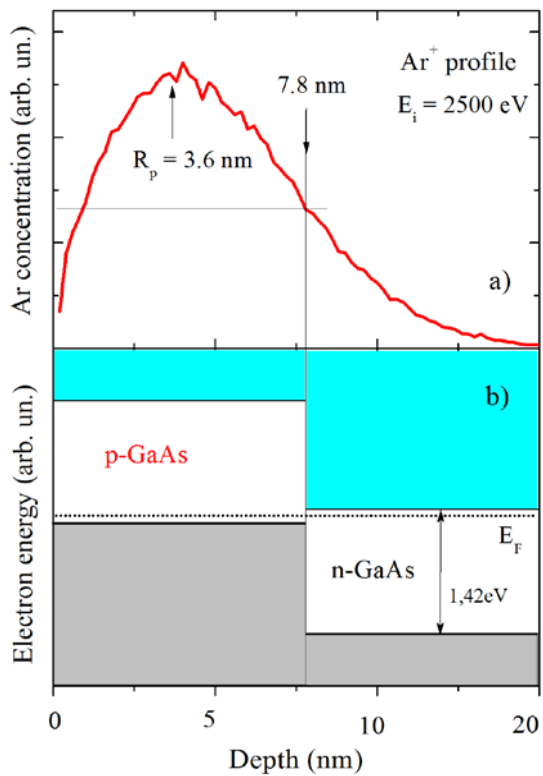

**Fig. 2.** (a) Depth distribution of the concentration of the thermalized Ar projectiles irradiating the n-GaAs (100) surface. (b) Band diagram of the p-GaAs nanolayer formed on the n-GaAs (100) surface by Ar$^+$ ion irradiation.

### III. RESULTS AND DISCUSSION

**A. Elemental and chemical composition of the initial and ion etched GaAs near-surface layers.**

Elemental composition of the GaAs initial surface and its modification due to Ar$^+$ ion bombardment was quantitatively studied in this work by measuring the photoelectron spectra in the relatively narrow energy range containing the As3d and Ga3d lines. Fig.1b shows the spectra measured at different photon energies in the binding energy scale. The spectra exhibit chemical shifts of the lines due to removal of the native oxide layer from the initial surface by Ar$^+$ ions and some redistribution of the line intensities evidencing modification of the elemental composition. The spectra were normalized to the area of the As3d line, that is to the atomic concentration of As atoms. Nevertheless, direct comparison of the Ga3d line intensities does not give the relative elemental composition and its modification due to Ar$^+$ etching since relative line intensities are shaped by the excitation probability depending on the photon energy. The elemental composition was obtained by normalizing the line intensities (areas) to the corresponding photoemission cross-sections [19]. The obtained relative concentrations of As and Ga atoms in the GaAs near-surface layer are given in Table I.

The spectra were measured at different photon energies to probe the near-surface layers of different thickness. The information depth is directly determined by the mean free path ($\lambda$) of normally ejected photoelectrons. The mean free path value for GaAs was taken from theoretical calculations [20]. To obtain the unknown data for GaAs oxides, the modified Bethe equation simple in form was used [21] for $Ga_2O_3$, $As_2O_3$ and GaAs. Then, the roughly estimates data were reduced to more reliable data of Ref. [20] for GaAs by comparing the estimates for GaAs with those obtained in Ref [20]. Possible error of the procedure was assumed to be small since the estimated photoelectron mean free path in oxides proved to be only ~10% longer than that estimated for pure GaAs. The previously known and obtained here mean free paths are given in Table 1. These data shows that the elemental content was probed in the range as far as $\lambda \sim 3$ nm, covering the surface and the bulk of studied material.

**Table I.** Relative concentrations of As and Ga atoms in the GaAs near-surface layer of thickness $\lambda$ before and after etching of the native oxide layer by Ar$^+$ ions of energy $E_i$ = 2500 eV.

| $h\nu$ eV | $\lambda$ nm | [As] | [Ga] | [Ga]/[As] |
|---|---|---|---|---|
| Pristine native oxide surface ||||| 
| 800 | 2.8 | 0.44 | 0.57 | 1.30 |
| 650 | 2.5 | 0.39 | 0.61 | 1.56 |
| Ar$^+$ etched surface ||||| 
| 650 | 2.5 | 0.46 | 0.54 | 1.24 |
| 150 | 1.0 | 0.45 | 0.55 | 1.22 |

Table I represents relative concentrations of As and Ga atoms in the GaAs near-surface layer of thickness $\lambda$ before and after Ar$^+$ etching of the native oxide layer which consists of a mixture of $Ga_2O_3$ and $As_2O_3$ oxides as it is shown below (see Fig.4a). The data on initial native oxide layer shows that oxidation of the GaAs surface results in strong enrichment of the layer by gallium. The Ga enrichment appeared to be more significant when approaching the surface, since the ratio [Ga]/[As] increases (1.3 → 1.6) when information depth $l \sim \lambda$ becomes smaller (2.8 → 2.5). The essential lack of $As_2O_3$ in thermal oxide has been explained in former studies by relatively high volatility of this oxide [22]. Indeed, the $As_2O_3$ melting point is 312.2$^o$C contrary to 1900$^o$C of $Ga_2O_3$. A different mechanism may be assumed for native oxidation at room temperature: Ga atoms are quickly oxidized due to higher affinity for oxygen and remain fixed at some positions while As atoms diffuse away.

Table I shows that, contrary to significant Ga-enrichment of the native oxide layer, ion irradiation induces a rather moderate Ga-enrichment of the near-surface layer: the Ga atomic concentration is higher than the As concentration only by a factor of ~ 1.2 at ion energy $E_i$ = 2500 eV, which is in qualitative agreement with the previously reported effect [12-14]. In this case, the enrichment mechanism is related to preferential sputtering of As atoms [12]. The data obtained at different photon energies ($h\nu$ = 650, 150 eV) demonstrates to a first approximation a homogeneous distribution of Ga atoms throughout the depth of about 2.5 nm. Though, Table I shows some depletion of the Ga concentration comparable to the measurement error in near-surface layers. The point important for physical properties of the irradiated layer is what is the chemical state of the excess Ga atoms. Analysis of the spectral line chemical shifts is presented in Sec. C.

## B. Determination of the conductivity type of the ion-treated GaAs surface.

Physical properties of pristine and treated GaAs near-surface layers were studied via photoemission valence band spectra characterizing the density of occupied states. Fig. 3 demonstrates the valence band spectra of a pristine native oxide (GaAs-ox) layer covering the surface of an *n*-GaAs wafer, as well as spectra from the same sample after removal of the native oxide and ion irradiation of the atomically clean GaAs surface (*p*-GaAs). The spectra of the ion-treated surface were measured relative to the Fermi level with ordinary energy resolution $\Delta E \sim 300$ meV (dotted curve) and high energy resolution $\Delta E \sim 150$ meV (solid curves). Comparison of the spectra measured with different resolutions shows that the spectrum tail above the Fermi level is caused by the finite energy resolution of the spectrometer. However, conventional extrapolation of the spectrum brink enables determination of the valence band edge position which, as the figure shows, proves to be ~ 0.14 eV below the Fermi level for the ion etched surface and ~ 1.0 eV for the initial oxidized surface. The closeness of the valence band edge of the ion-treated surface to the Fermi level evidences for occurrence of a *p*-type of the near-surface layer after removing the native oxide and ion irradiation of the *n*-GaAs near- surface layer. Thus ion irradiation has been revealed to change the GaAs semiconductor type from *n* to *p*, resulting in creation of the *p-n* junction whose scheme is presented in Fig. 2b. This process manifests itself in shifting the valence band edge by ~ 1 eV (Fig. 3.)

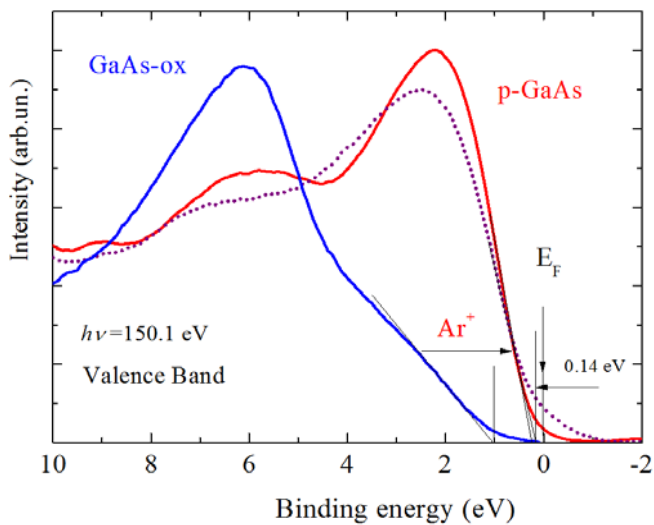

**Fig. 3.** Valence band (VB) photoelectron spectrum of a pristine n-GaAs(100) wafer with a layer of native oxide (GaAs-ox) and the p-GaAs nanolayer formed on the n-GaAs(100) surface by Ar$^+$ ion irradiation (p-GaAs). The spectra were measured with ordinary energy resolution $\Delta E \sim 300$ meV (dotted curve) and a higher energy resolution $\Delta E \sim 150$ meV (solid curves).

The VB photoelectrons measured at photon energy $h\nu = 150$ eV are characterized by short mean free path $\lambda = 1$ nm. This fact allows the conclusion that the *p*-layer thickness exceeds 1 nm. However the VB spectra do not provide more accurate information and leave open the question about the thickness of the *p*-type layer. The answer to this question has been obtained by studying the *p-n* junction via the core-level spectra represented in Sec. C.

## C. Studying the *p-n* junction via core-level spectra.

The chemical state of atoms in the GaAs near-surface layer was characterized by core-level photoelectron spectra. These spectra also confirm the revealed conversion of the conductivity type of the *n*-GaAs near-surface layer under ion irradiation. Fig. 4a shows Ga3d photoelectron spectra of the sample before (Ga$_2$O$_3$) and after (*p*-GaAs) ion etching. The spectra were measured at the photon energy $hv$ = 150 providing information from the bulk as deep as $l \sim 2\lambda = 2$ nm. The spectrum of the initial surface consists of a broad peak centered at the energy E$_B$ = 20.5 eV corresponding to Ga$_2$O$_3$ according to the data of Table II. This observation evidences a decay of the GaAs near-surface layer in oxidation process resulting in formation of a mixture of elementary oxides (Ga$_2$O$_3$ , As$_2$O$_3$). The Ga3d line "*p*-GaAs" in Fig. 4a was detected after complete removal of the native oxide layer, which was also controlled by the oxygen O1s photoelectron line (Fig. 1a). The Ga3d line consists of an unresolved doublet (3d$_{3/2,1/2}$). The averaged binding energy of this line almost fully coincides with our former data [23] and proves to be very close to the previously published data (Table II) obtained for the GaAs surfaces prepared in completely different ways resulting in either crystalline or amorphous structures [6,7]. Approximate equality of the binding energies evidences the proximity of the short-range order in crystalline GaAs and GaAs made amorphous by ion or chemical etching. The spectrum of the modified layer enriched with gallium may contain an essential unresolved contribution of a metallic phase at E$_B$ = 18.6 eV. Thus, the considered data evidence formation by ion etching a well-defined state of the atomically clean GaAs surface and near-surface layer enriched with Ga atoms more or less homogeneously (Table I).

**Table II.** Binding energies (E$_B$) of Ga3d core-electron in different materials averaged over the data of Ref. [6,7] and of this work. The E$_B$ values are counted from the Fermi-level.

|  | Ga3d E$_B$ , eV | Ref. |
|---|---|---|
| Ga-metal | 18.6 | 6,7 |
| Ga-metal | 18.6 | this work |
| GaAs-cath.amorph. | 19.0 | 6,7 |
| GaAs-cleav. | 19.3 | 6,7 |
| GaAs-chem. etch. | 19.3 | 6,7 |
| GaAs-Ar+ etch. | 19.4 | 23 |
| p-GaAs-Ar+ etch. | 19.3-19.35 | this work |
| n-GaAs. | 18.0 | this work |
| Ga$_2$O$_3$ | 20.3 – 21.0 | 6,7 |
| Ga$_2$O$_3$ | 20.5 | this work |

In discussing the data on the binding energies in GaAs, the only assumption should be made that all the E$_B$ data in Table. II refer to the same type of the semiconductor conductivity (*n* or *p*). Otherwise, the data spread should be of about the GaAs bandgap width ($\Delta E_g$ = 1.42 eV) because the binding energy counted from the Fermi-level increases by the Fermi-level shift in the bandgap ($\Delta E_F \sim \Delta E_g$) in turning from *p*- to *n*-type. We show here that the Ga3d binding energies for GaAs obtained in this and other works ( see Table II) refer to the *p*-type GaAs. This fact follows from the valence band (Fig. 3) and Ga3d core-level (Fig. 4,5) spectra discussed below.

Fig. 4b presents a Ga3d photoemission spectrum of the ion-treated GaAs surface measured at essentially higher photon energy $hv$ = 650 eV (and larger pass energy). In this case, the information depth ($l \sim 2\lambda$ = 5 nm) was comparable to the thickness of the irradiated layer and expected to be sufficient to observe the unirradiated *n*-GaAs layer underneath. As mentioned above, the Ga3d line of the *n*-type layer should be shifted towards higher binding energies by ~ $\Delta E_g$ = 1.42 eV. Indeed, the spectrum in Fig. 4b may contain the expected line at E$_B$ = 20.7 eV in the high-energy tail of the main line. Unfortunately a reliable conclusion cannot be made from the spectrum presented in Fig 4b since the instrumental function in the mode with a larger pass energy has proved to be a bit specific. For this reason, additional data obtained using another spectrometer (CLAM-4) with better instrumental function and higher resolution were involved into the research.

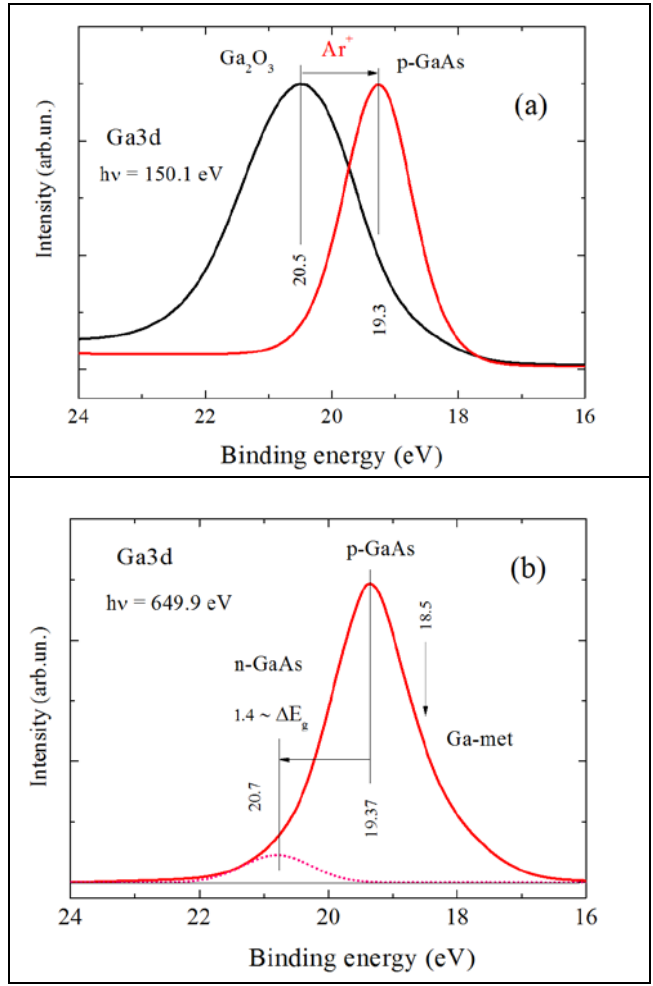

**Fig. 4.** Ga3d photoelectron spectra of a pristine *n*-GaAs(100) wafer covered by a layer of native oxide (Ga$_2$O$_3$) and of the same sample after removal of the oxide layer by the beam of Ar$^+$ ions with energy E$_i$ = 2500 eV (*p*-GaAs): **(a)** spectra measured at the photon energy $h\nu$ = 150 eV and **(b)** the spectrum measured at the photon energy $h\nu$ = 650 eV. A possible contribution of initial *n*-GaAs bulk layer underneath the irradiated surface layer (dotted curve) is shown in panel (b).

Fig. 5 shows Ga3d photoemission spectra of the GaAs surface treated by Ar$^+$ ions of energy E$_i$ = 1500 eV and measured with the spectrometer CLAM-4 in one and the same mode at two photon energies $h\nu$ = 150 and 350 eV characterizing thin and relatively thick layers of material. To completely exclude possible influence of negligible oxidation in high vacuum during measurements on the spectral shape, the spectrum presumably containing the sought-for effect at higher photon energy was measured first. Fig.5a shows also a difference between the spectra. This difference reveals a small but statistically reliable contribution of the top of the *n*-GaAs bulk underneath the modified *p*-GaAs layer. The binding energy of the revealed contribution (E$_B$ = 20.15 eV) proved to be less than the maximal value (E$_B$ = 20.7 eV) possible for *n*-GaAs because of an insufficient photoelectron mean free path ($\lambda$ ~1.8 nm) restricting the probing depth ($l$ ~ 2$\lambda$ = 3.6 nm) to the area lying between the *p*-layer and *n*-bulk, namey, the upper part of the *p*-GaAs/*n*-GaAs interface. The revealed *n*-GaAs contribution is also confirmed by second derivatives of the initial spectra (Fig. 5b). In the differentiated spectrum measured at $h\nu$ = 350 eV, this contribution looks like a negative peak at binding energy E$_B$ = 20.2 eV. This negative peak reduces the intensity of the positive high-energy part of the main *p*-GaAs line. Using the revealed contribution of the top of the *n*-GaAs bulk into the Ga3d spectrum, it is possible to estimate the thickness of the *p*-GaAs layer as $d = \lambda \ln(I_p/I_n +1)$, where $I_p$ and $I_n$ are the *p*- and *n*-contributions (areas) to the Ga3d line intensity. The estimation gives value $d$ = 5.5 nm which proved to be equal to the calculated thickness $\langle x \rangle \sim 2R_p = 5.4$ nm of the layer irradiated by the 1500-eV ions. A replicate experiment gave a close value: $d = 4.7$ nm.

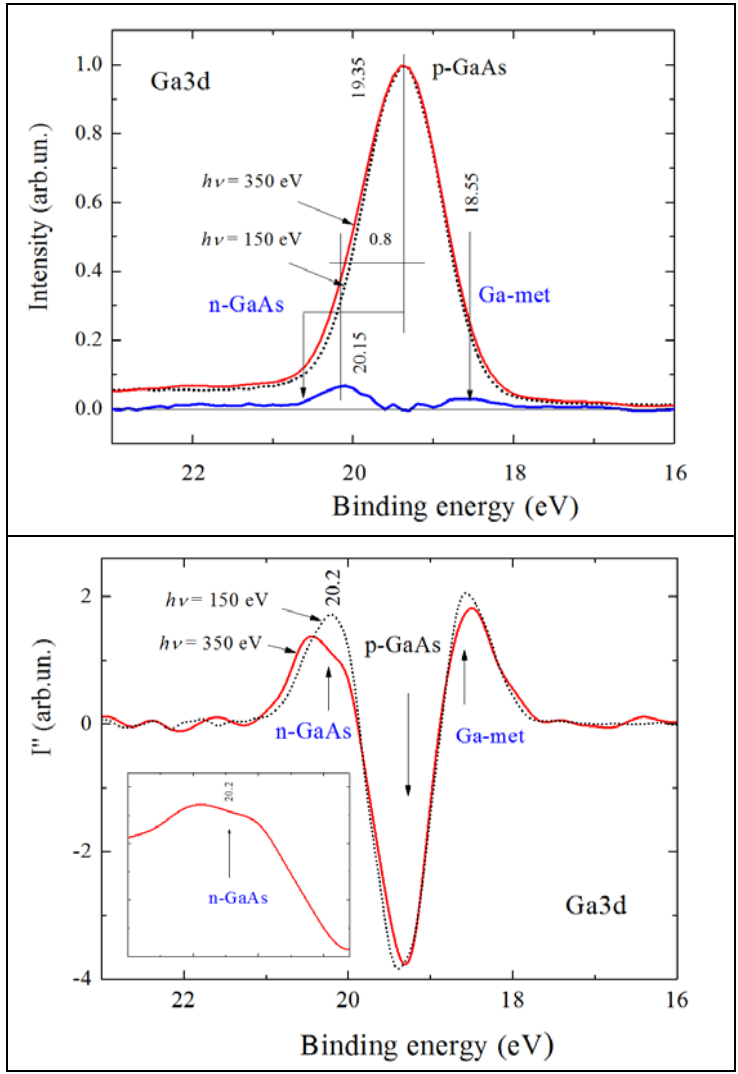

**Fig. 5.** (a) Ga3d photoelectron spectra of the sample n-GaAs irradiated by the 1500-eV Ar$^+$ ion beam and measured at the CLAM-4 (VG) station with photon energies $h\nu = 150$ eV and $h\nu = 350$ eV. (b) Second derivative of the spectrum measured at photon energy $h\nu = 350$ eV and presented in panel (a).

As mentioned above, Ar$^+$ ion etching of the GaAs near-surface layers results in their amorphization and enrichment of the irradiated layer with gallium due to preferable sputtering of arsenic atoms. Therefore, transformation of the conductivity type in the irradiated layer should be naturally associated with these processes. At the Ar$^+$ ion energy $E_i = 2500$ eV, the gallium to arsenic ratio reaches the value [Ga]/[As] =1.2 (see Table I). So, approximately 20 % of Ga atoms can create interstitial and antisite defects as well as segregate into metallic clusters. Indeed, Fig 5 demonstrates the presence of metallic Ga at binding energy $E_B = 18.55$ eV when the photon energy increases from 150 to 350 eV and, thus, the thickness of the layer accessible for studying increases. Enrichment with metallic Ga of deeper layers can be explained by more rapid diffusion of Ga atoms away from the surface and near-surface layers even at room temperature. Analysis of the Ga3d line showed that the metallic phase contains only a part of the excess Ga atoms. An essential part of Ga atoms can occupy the As positions and create antisite defects imparting to the material the *p*-type conductivity. On the other hand, the initial concentration of *n*-doping atoms should decrease under the ion bombardment due to their excitation into interstitial position followed by outward diffusing.

Thereby, bombardment of the n-GaAs surface was shown to change the conductivity type in the near-surface layer throughout the irradiated layer thickness approximately equal to double projected range. For ion energy $E_i = 1500$ eV, it is equal to $d = 5.5$ nm ~ $<x>$ ~ $2R_p$. As a result, a *p-n* junction arises. Fig.2b shows the scheme of the *p-n* junction with an abrupt interface for $Ar^+$ ion energy $E_i = 2500$ eV. In fact, the *p-n* junction width should be comparable to the area where the concentration of the created *p*-type defects becomes higher than that of *n*-doping atoms. In any case, the junction width should be much less than the tail distribution of the thermalized Ar atoms represented in Fig.2a. Evidently, the obtained thin nanostructure can be cleaned from relatively loosely coupled interstitial and metallic Ga atoms by heating.

## III. CONCLUSION

In conclusion, the change of the conductivity type from *n* to *p* has been revealed in the well defined *n*-GaAs layer irradiated by an $Ar^+$ ion beam. The *p*-layer thickness was shown to virtually coincide with the double projected range which can be easily calculated by SRIM code. For example, the *p*-layer thickness in the studied ion energy range $E_i = 1.5 - 2.5$ keV turns out to the range from 5.5 to 7.2 nm. As a result, the *p-n* junction was created on the GaAs-surface by the ion beam. The effect manifests itself both in shifting the core-levels to higher binding energies by the value of about the bandgap width and in shifting the valence band edge to lower binding energies as far as the edge abutment to the Fermi-level. Transformation of the conductivity type was assumed to be connected with preferable sputtering of arsenic atoms and enrichment of the irradiated layer with gallium followed by creation of the Ga antisite *p*-type centers. The revealed effect can be used to form *p-n* nanojunctions on the GaAs surface directly by an $Ar^+$ ion beam with the lateral resolution determined by the ion beam diameter.

## ACKNOWLEDGMENT

The research was partially supported by the Russian Foundation for Basic Research, Project N 16-02-00665-a. The authors thank HZB for the allocation of synchrotron radiation beamtime and the Russian German Laboratory at BESSY II Helmholtz Zentrum Berlin for the support in the synchrotron radiation study.